\documentclass[10pt]{article}

\usepackage[a4paper,textwidth=18cm,textheight=24cm,top=2.85cm, bottom=2.85cm, left=1.5cm, right=1.5cm]{geometry}
\pdfoutput=1
\usepackage{scicite}
\usepackage{hyperref}
\usepackage{icdp2009}
\usepackage{lmodern}
\usepackage{graphicx, amsmath, amsfonts}
\usepackage{times}

\makeatletter
\long\def\@makecaption#1#2{%
\vskip\abovecaptionskip
\sbox\@tempboxa{#1. #2}%
\ifdim \wd\@tempboxa >\hsize
#1. #2\par
\else
\global \@minipagefalse
\hb@xt@\hsize{\box\@tempboxa\hfil}%
\fi
\vskip\belowcaptionskip}
\makeatother

\newcommand{\fref}[1]{Figure~(\ref{#1})}
\newcommand{\eref}[1]{Equation~\eqref{#1}}
\newcommand{\xpar}{\\\par\noindent}
\newcommand{\ypar}{\par\noindent}
\begin{document}
\noindent

\bibliographystyle{ieeetr}

\title{Optimal Background Correction in Double Quantum Coherence Electron Spin Resonance Spectroscopy for Accurate Data Analysis}

\authorname{Aritro Sinha Roy}
\authoraddr{\small Department of Chemistry \& Chemical Biology \\ \small Cornell University, Ithaca, New York, USA}


\maketitle

\abstract
Electron spin resonance (ESR) pulsed dipolar spectroscopy (PDS) is used in protein 3D structure determination. However, the accuracy of the signal analysis depends heavily on the background correction process. In this work, we derive the functional forms of double quantum coherence (DQC) ESR signal in typical frozen samples of micro-molar concentration, quantifying both the intramolecular and the background contributions. This is a draft manuscript, minor updates will be made once published.


\section*{Introduction}
ESR pulsed dipolar spectroscopic techniques, coupled with site directed spin labeling (SDSL) method \cite{SDSL_Hub1,SDSL_Hub2} have emerged as a set of sophisticated tools in measuring distance distributions between specific pairs of protein residues to study its 3D structure \cite{Freed_Science,DEER_Rev}. However, extraction of the distance distribution from PDS signals is an ill-posed problem \cite{ill_posed_1,ill_posed_2} and a small perturbation in the signal can introduce a large error in the solution \cite{ill_posed_3, opt_tkn}. Hence, removing the intermolecular signal contribution or \emph{background correction} of the signal becomes a critical step in the analysis \cite{conc2,dist_measure_jeschke}. In this work, we focus on the double quantum coherence (DQC) PDS technique \cite{Freed_Science, DQC_Theo1, MQC_Theo1, Acert_DQC1}. Previously, the background signal of DQC have been suggested based on experimental and/or empirical studies \cite{MQC_Theo1,DQC_Theo1}. Given that there remains some doubt regarding the background contribution, often additional experiments are conducted and/or polynomial curve fittings are employed for background correction \cite{dq_sft_molecule, pds_prisner}, limiting both the applicability of the methods, and the consistency of the signal analysis. Therefore, we derive the analytical expression of DQC signal, originating from a frozen sample of $N$ spin labeled molecules, and obtain the functional forms of the signal by averaging over the orientational parameters. Additionally, the effect of finite size of the spin labeled protein molecules on the background signal is discussed, since they should not be treated as point particles \cite{Protein_Size}.
\section*{Theoretical Background}
\subsection*{Density operator evolution}
\begin{figure}[h]
\begin{center}
    \includegraphics[width=0.8\linewidth]{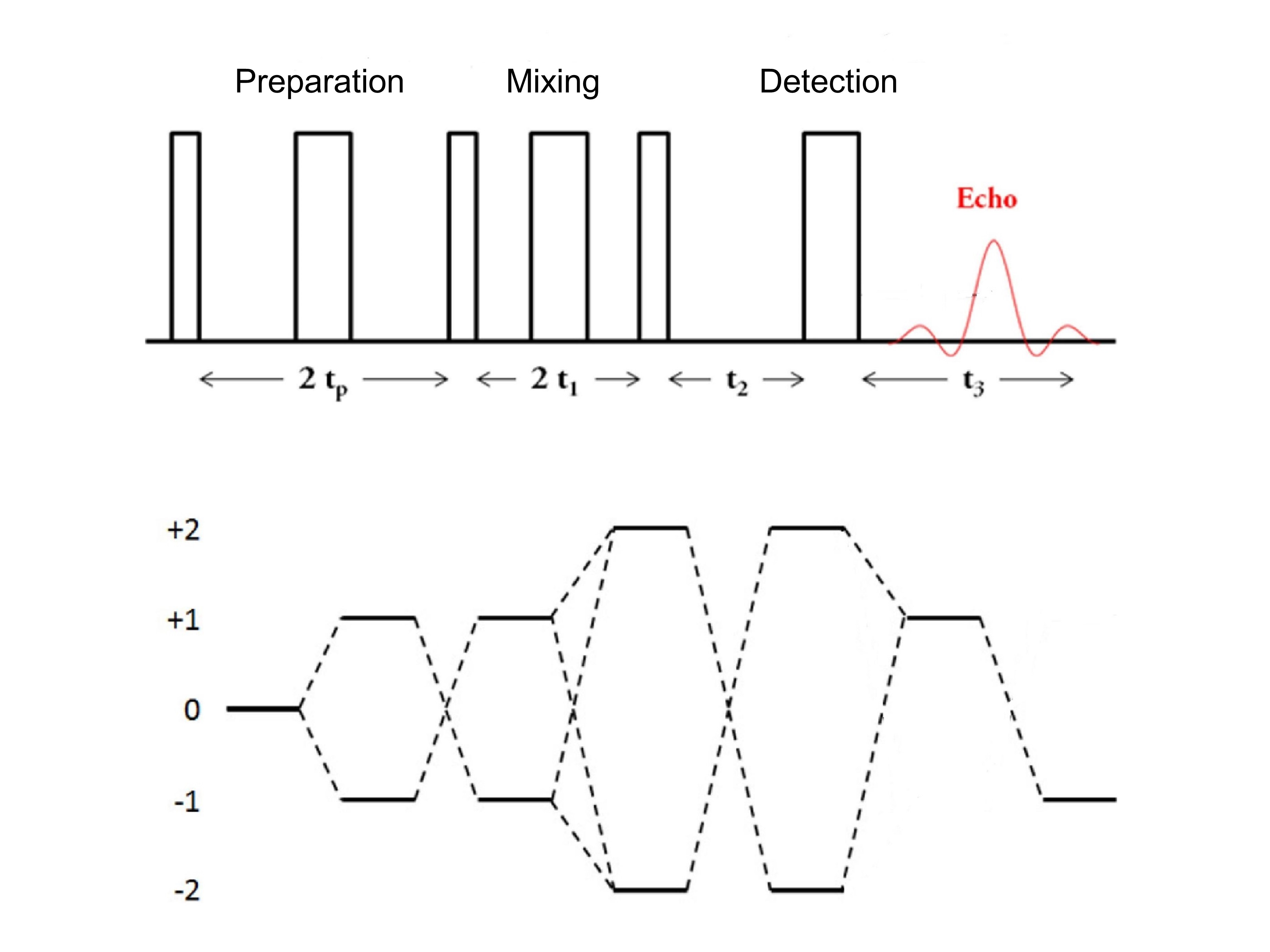}
    \caption{Pulse sequence of DQC and the associated coherence pathway. The $\pi$ pulses are represented by rectangular bars with twice the width of that of the $\pi/2$ pulses. DQC signal is plotted against $t_\xi = t_2 - t_p$, keeping the dipolar evolution time, $t_m = t_2 + t_p$ constant and setting $t_3 = t_2$.}
  \label{fig:pulse_seq}
\end{center}
\end{figure}

In deriving the analytical expressions, we use the following approximations: (1) the secular spin hamiltonian is used \cite{ESR_Methods}, (2) the pulses are considered to be ideal, and (3) we hypothesize the spins as $S=1/2$ point particles.
\xpar In 6-pulse DQC, the $\pi/2-t_p-\pi-t_p-\pi/2$ sequence selectively generates the double quantum coherence, while the inversion sandwich, $t_1-\pi-t_1$ is used for the evolution. The last $\pi/2$-pulse in the sequence converts the DQC to anti-phase single quantum coherence (SQC), which evolves into in-phase SQC signal by the $t_2-\pi-t_2$ inversion sandwich. The pulse sequence and the selected coherence pathway are shown in \fref{fig:pulse_seq}.
The evolution of the density operator under the influence of a hamiltonian, $H$ is given by the time-independent Liouville-Von Neumann equation as \cite{Slichter}
\begin{equation}
\rho(t\,+\,\delta t) = e^{-\,i\,H\,\delta t}\,\rho(t)\,e^{\,i\,H\,\delta t}
\end{equation}
The secular hamiltonian ($H$), the propagators of free evolution ($Q$), and the microwave pulses ($R$) are given by
\begin{equation}
\begin{split}
& H = \sum_j^N{\Omega_j\,S_{jz}} + \sum_{j<k}{a_{jk}\,S_{jz}\,S_{kz}} \\
& Q(t) = \exp{\left(-i\,H\,t\right)} \\
& R_\phi(\beta) = \exp{\left(-i\,\beta\,\sum{S_x\,\cos{\phi}\,+\,S_y\,\sin{\phi}}\right)}\\
\label{eqn:propagators}
\end{split}
\end{equation}
where $\Omega_j$, $a_{jk}$, $\beta$, and $\phi$ are the resonance offset of the $j^{th}$ spin, the dipolar coupling constant between the $\{j,\,k\}$ spin pair, the flip angle and the phase of the pulse, respectively, and $S_{x,y}$ corresponds to the $S=1/2$ angular momentum operator. Accounting for the fact that only a fraction of the total spins, $\delta$ are flipped by a finite inversion pulse, we note
\begin{equation}
R_\phi(\pi)\,S_j\,R_\phi(\pi)^\dagger \rightarrow (1-\delta)\,S_j + \delta\,R_\phi(\pi)\,S_j\,R_\phi(\pi)^\dagger
\label{eqn:fpulse_sandwich}
\end{equation}
Finally, we define the inversion sandwich operator as
\begin{equation}
\Xi(t, \phi) \equiv Q(t)\,R_\phi(\pi)\,Q(t)
\end{equation}
The evolution of the density operator in the 6-pulse experiment can be summarized as
\begin{equation}
\begin{split}
& \rho_0 \propto S_{1z} \\
& \rho_1 = \sum_{\phi_1}R_{\phi_1}(\pi/2)\,\Xi(t_p,\phi_1)\,R_{\phi_1}(\pi/2)\,\rho_0 \\
& \mkern 36mu R_{\phi_1}(\pi/2)^\dagger\,\Xi(t_p,\phi_1)^\dagger\,R_{\phi_1}(\pi/2)^\dagger \\
& \rho_2 = \sum_{\phi_2}\Xi(t_1,\phi_2)\,\rho_1\,\Xi(t_1,\phi_2)^\dagger \\
& \rho_2^+ = \sum_{\phi_3}R_{\phi_3}(\pi/2)\,\rho_2\,R_{\phi_3}(\pi/2)^\dagger \\
& \rho_3 = \sum_{\phi_4}\Xi(t_2, \phi_4)\,\rho_2^+\,\Xi(t_2, \phi_4)^\dagger \\
\end{split}
\label{eqn:dqc_evol_summary}
\end{equation}
It should be noted that we consider the equilibrium density operator, $\rho_0$ to be proportional to $S_{1z}$, since it is sufficient to calculate the interaction of spin-1 with the rest of the spins in deriving the average DQC signal.
\section*{Derivation of the DQC Functional Form}
\subsection*{Isolated pair of spins}
Following the scheme shown in \eref{eqn:dqc_evol_summary} and using a Mathematica spin rotation package\footnote{The original package is created by John Marohn, Department of Chemistry \& Chemical Biology, Cornell University}, the density operator for a pair of $S=1/2$ spin probes after the second $\pi/2$-pulse is given by
\begin{equation}
\rho_1 = (-1+\delta+\delta^2)\,s_{a}^{t_p}\,\big(S_{1x}\,S_{2y}+S_{1y}\,S_{2x}\big)
\label{eqn:dqc_pair_prep}
\end{equation}
where $s_{a}^{t_p} = \sin{\left(a\,t_p\right)}$, and $a$ is the dipolar coupling constant of the interaction between the pair of spin probes. At the end of the mixing period, the density operator evolves to
\begin{equation}
\rho_2 = -(-1+\delta+\delta^2)^2\,s_{a}^{t_p}\,\big(S_{1x}\,S_{2y}+S_{1y}\,S_{2x}\big)
\label{eqn:dqc_pair_mix}
\end{equation}
Finally, the isolated $S=1/2$ pair intramolecular DQC signal expression is given by
\begin{equation}
V_{0,dqc} = (i/8)\,\delta^2\,(-1+\delta+\delta^2)^2\,s_a^{t_2}\,s_a^{t_p}
\label{eqn:dqc_pair_signal}
\end{equation}
We have averaged the signal over the relevant phase cycles, and utilized $Tr(S_{j-}\cdot{S_{j+}})/Tr(I_4) = i/2$ in writing \eref{eqn:dqc_pair_signal}, where $I_n$ is the unit matrix of order $n$, which equals 4 in the direct product operator basis of a pair of $S=1/2$ particles.
\xpar We note that $a = \omega_0\,(1-3\,\cos{\theta}^2)/r^3$, where $r$ and $\theta$ are the amplitude of the vector joining the spin pairs, and the angle it makes with the direction of the external magnetic field. The powder averaged signal in the solid state is given by
\begin{equation}
\begin{split}
\langle V_{0,dqc}\rangle_\theta &= \int_\theta\sin{\theta}d\theta\,V_{0,dqc} \\
&= A_1\,\big(B(\omega_0, t_\xi) - B(\omega_0, t_m)\big) \\
\label{eqn:dqc_kernel}
\end{split}
\end{equation}
We denote the Fresnel cosine, and sine integrals as $fc$ and $fs$ in defining $A_1$ and $B$ as
\begin{equation}
\begin{split}
& A_1 = i\,\sqrt{\pi}\,\delta^2\,(-1+\delta+\delta^2)^2/16\sqrt{6} \\ & B(\omega,t) = \frac{c_\omega^t\,fc(\sqrt{6\,\omega\,t/\pi}) + s_\omega^t\,fs(\sqrt{6\,\omega\,t/\pi})}{\sqrt{\omega\,t}}\\
\end{split}
\end{equation}
\subsection*{N-spin system}
In case of an $N$-spin system, we use ideal pulses in evaluating the density operators, and introduce the effect of the finite pulse by substituting the sample concentration with an \emph{effective} concentration in the final expression. We have evaluated the density operators in the 6-pulse DQC explicitly for systems with 2 to 8 spins, and applied the method of induction in deriving the density operator at the end of the preparation period, $\rho_1$ as
\begin{equation}
\begin{split}
\rho_1 =& \,\sum_{n}^{N_\textsc{odd}}\sum_{\substack{j\subset{\{N\}} \\ |j|=2\,n-1 \\ 1\notin{j}}}(-1)^{(n-1)/2}\,2^{n-1}\,\prod_{j}^{n}{s_{a_{1j}}^{t_p}} \\ &
\prod_{k\colon\{N\}-j\cup{1}}{c_{a_{1k}}^{t_p}}\,(S_{1x}\,\prod_{j}^{n}{S_{jy}}+S_{1y}\,\prod_{j}^{n}{S_{jx}}) \\ & + \,\sum_{n}^{N_\textsc{even}}\sum_{\substack{j\subset{\{N\}} \\ |j|=2\,n \\ 1\notin{j}}}(-1)^{(n-2)/2}\,2^{n-1}\,\prod_{j}^{n}{s_{a_{1j}}^{t_p}} \\ &\prod_{k\colon \{N\}-j\cup{i}}{c_{a_{1k}}^{t_p}}\,(S_{1z}\,\prod_{j}^{n}{S_{jx}}-S_{1z}\,\prod_{j}{S_{jy}})
\end{split}
\label{eqn:ndqc_prep}
\end{equation}
where the notation $j\subset{\{N\}} \colon |j| = 2\,n-1$ denotes all the subsets of $\{N\}$ of size $2\,n-1$.
\xpar Going into the mixing period, it should be noted that in an $N$-spin system, both multi-spin DQC and multiple-quantum coherence (MQC) (order$\,>\,$2) are produced. However, if the value of $t_1$ in the mixing period is kept constant, and within its usual range of 10-20 ns \cite{DQC_Theo1}, the mixing period can be reduced to the application of a $\pi$-pulse. In other words, considering $t_1\to 0$, we obtain
\begin{equation}
\rho_2 = -\,\rho_1
\label{eqn:dqc_mixing}
\end{equation}
After the last $\pi/2$-pulse, only anti-phase SQC signal is selected, which evolves into detectable in-phase SQC at the end of the detection period and the corresponding density operator is given by
\begin{equation}
\begin{split}
\rho_2^+ =& -(1/2)\,\bigg[(S_{1+}\,S_{jz} + S_{j+}\,S_{1z})\,s_{a_{1j}}^{t_p}\,\prod{c_{a_{1k}}^{t_p}} \\
& +\,\sum_{n=3}^{N_\textsc{odd}}{i^{n-1}\,2^{n-1}\,S_{1+}\,\prod_{j\ne{1}}{S_{jz}\,s_{a_{1j}}^{t_p}}\,\prod{c_{a_{1k}}^{t_p}}} \\
& -\,\sum_{n=2}^{N_\textsc{even}}{i^{n-2}\,2^{n-1}\,S_{1+}\,\prod_{j\ne{1}}{S_{jz}\,s_{a_{1j}}^{t_p}}\,\prod{c_{a_{1k}}^{t_p}}}\bigg] \\
\end{split}
\label{eqn:dqc_detA}
\end{equation}
The conversions of multi-spin anti-phase SQC (+1) to in-phase SQC (-1) by an ideal inversion sandwich is given by
\begin{equation}
\begin{split}
& S_{p+}\prod_{m=1}^{N-1}{S_{jz}} \rightarrow \left(-\frac{i}{2}\right)^m\,\Delta^{m+1}\,S_{p-} \\ & \mkern 54mu \prod_j{\sin{(a_{pj}\,t_2)}}\,\prod_{k\ne p,j}{\cos{(a_{pk}\,t_2)}} \\
\end{split}
\label{eqn:ndqc_fast_eval}
\end{equation}
Using the expressions in \eref{eqn:dqc_detA}, and \eref{eqn:ndqc_fast_eval}, the $N$-spin DQC signal can be written as
\begin{equation}
\begin{split}
S(t_p, t_2) =& (i/8)\,
\bigg[(N-1)\,s_{a_{12}}^{t_p}\,s_{a_{12}}^{t_2} \\
& \prod_{k\ne{1,2}}{c_{a_{1k}}^{t_p}\,c_{a_{1k}}^{t_2}}\,+ \,\sum_{n=1}^{N-1}\,(-1)^{n-1}\,\binom{N-1}{n}\,\\
& \prod_{j=2}^{n+1}{s_{a_{1j}}^{t_p}\,s_{a_{1j}}^{t_2}}\,\prod_{k=n+2}^{N}{c_{a_{1k}}^{t_p}\,c_{a_{1k}}^{t_2}}\bigg] \\
\end{split}
\label{eqn:ndqc_detectionB}
\end{equation}
We introduce an additional spin-1$^\prime$ in the system, which along with the spin-1 at the origin constitutes the intramolecular pair and by replacing $N$ with $N+1$, we rewrite \eref{eqn:ndqc_detectionB} as
\begin{equation}
S(t_p, t_2) = -(i/8)\,\sum_{n=1}^N{\binom{N}{n}\,(-x)^n\,y^{N-n}}
\label{eqn:dqc_tot}
\end{equation}
In writing \eref{eqn:dqc_tot}, we have assumed that (i) the system has $N$ intermolecular spins and an intramolecular pair, (ii) $x = s_a^{t_p}\,s_a^{t_2}$, (iv) $y = c_a^{t_p}\,c_a^{t_2}$. We recall that the isolated pair signal, $v_0 \propto (i/8)\,s_{a_{11^\prime}}^{t_p}\,s_{a_{11^\prime}}^{t_2}$, and collect the terms containing $v_0$ to express the intramolecular DQC signal as
\begin{equation}
\begin{split}
V_{ir} &= \,\delta^2\,(-1+\delta+\delta^2)^2\,v_0\,\sum_{n=0}^{N-1}{\binom{N-1}{n}\,(-x)^n\,y^{N-n}} \\
&= \,\delta^2\,(-1+\delta+\delta^2)^2\,v_0\,(y-x)^{N-1}
\label{eqn:dqc_nintra1}
\end{split}
\end{equation}
With $N\rightarrow\infty$, the intermolecular contribution becomes
\begin{equation}
\begin{split}
V_{it} &= -\,(i/8)\,\sum_{n=1}^{N}{\binom{N}{n}\,(-x)^n\,y^{N-n}} \\
&= -\,(i/8)\,\bigg[\sum_{n=0}^{N}{\binom{N}{n}\,(-x)^n\,y^{N-n}}\,-\,y^N\bigg] \\
&= -\,(i/8)\,\big[(y-x)^N - y^N\big]
\label{eqn:dqc_ninter1}
\end{split}
\end{equation}
\textbf{Case-I ($0\le r\le{\infty}$):} In the ideal case, we consider the sample to be homogeneously distributed particles, and obtain the average signal expression by evaluating the integrals, $I_1 = \langle(y-x)\rangle_V$, and $I_2 = \langle y\rangle_V$. Recognizing that $(y-x) = \big\langle \cos{\left(a(r,\theta,\phi)\,t_m\right)} \big\rangle_V$, $I_1$ is evaluated as follows \cite{abragam_1961}
\begin{equation}
\begin{split}
I_1 &= \frac{1}{V}\,\int_\phi{d\phi\,\int_\theta{\sin{\theta}\,d\theta\,\int_r{r^2\,\cos{\left(b\,t_m/r^3\right)}dr}}} \\
& \mkern-18mu \text{considering } q = 1/r^3,\,\, b = \left|\omega\,(1-3\,\cos{\theta}^2)\right|\,t_m \\
& \mkern-18mu \text{as } r\to 0,\,q\to\infty \text{ and as } r\to\infty,\,q\to 0 \\
&= \frac{1}{3\,V}\,\int_\phi{d\phi\,\int_\theta{\sin{\theta}\,d\theta\,\int_{0}^{\infty}{\cos{(b\,q)}/q^2\,dq}}} \\
& \mkern-18mu \text{writing } \int_V{dV} \text{ as } \int_\phi{d\phi\,\int_\theta{\sin{\theta}\,d\theta\,\int_q{dq}}} \\
&= 1-\frac{1}{3\,V}\,\int_V{\Big(1-\cos{(b\,q)}\Big)/q^2\,dV} \\
& \mkern-18mu \text{changing the limit of q to } \{-\infty, +\infty\} \\
&= 1-\frac{1}{6\,V}\,\int_V{\Big(1-\cos{(b\,q)}\Big)/q^2\,dV} \\
&= 1\,-\,\frac{8\,\epsilon\,\pi^2\,C\,\omega\,t_m}{9\,\sqrt{3}\,N} \\
\end{split}
\label{eqn:integral_1A}
\end{equation}
where the spin labeled protein concentration, $C = N/V$ and $\epsilon = \delta\,(-1+\delta+\delta^2)$ corresponds to the efficiency of the inversion pulses and its value is derived from the probability factor in the isolated pair signal expression. The square root is used given that we are using $N$ intermolecular spins instead of $N$ pairs of spin labeled molecules. In the limit of $N\to\infty$, $(y-x)^{N-1}$ equals $\exp{\left(-8\,\epsilon\,\pi^2\,C\,\omega\,t_m/9\,\sqrt{3}\right)}$.
\xpar Next, $I_2$ is evaluated as follows
\begin{equation}
\begin{split}
I_2 &= \big\langle \left(\cos{\left(a\,t_m\right)}\,+\,\cos{\left(a\,t_\xi\right)}\right)/2 \big\rangle_V \\
&= 1-\frac{1}{6\,V}\,\int_V{\Big(1-\cos{(b_m\,q)}\,-\,\cos{(b_\xi\,q)}\Big)/q^2\,dV} \\
&= 1-\frac{4\,\epsilon\,\pi^2\,C\,\omega\,(t_m + t_\xi)}{9\,\sqrt{3}\,N} \\
\end{split}
\label{eqn:integral_2A}
\end{equation}
As $N\to\infty$, $I_2^{N}$ becomes $\exp{\left(-4\,\epsilon\,\pi^2\,C\,\omega\,(t_m + t_\xi)/9\,\sqrt{3}\right)}$. Combining the results, the intra and intermolecular 6-pulse DQC signal expressions are given by
\begin{equation}
\begin{split}
V_{ir} =& \,V_0\,\exp{\left(-2\,
\epsilon\,\eta\,t_m\right)} \\
V_{it} =& -\frac{i}{8}\,\bigg[\exp{\left(-2\,\epsilon\,\eta\,t_m\right)}\,-\,\exp{\left(-\epsilon\,\eta\,(t_m + t_\xi)\right)}\bigg] \\
\label{eqn:dq_tot_case1}
\end{split}
\end{equation}
where $\eta = 4\,\pi^2\,C\,\omega/9\,\sqrt{3}$. For organic radical pairs, $\omega$ equals $2\,\pi\,52.02\times10^{-21}\,s^{-1}\,m^3$ and $\eta$ becomes $8.278\,C\,\times\,10^{-4}\,\mu{s}^{-1}$, $C$ being expressed in $\mu{M}$.
\xpar \textbf{Case-II ($r_{min}\le r\le{\infty}$):} Inclusion of the effect of finite sized spin labeled proteins is achieved by assigning an empty spherical volume of radius $r_{min}$ around spin-1 at the origin, and we rewrite $I_{1}$ as
\begin{equation}
\begin{split}
I_1 &= 1-\frac{2\,\pi}{V}\,\int_\theta{\sin{\theta}\,d\theta\,\int_{r_{min}}^{\infty}{r^2\,\left(1-\cos{\left(b/r^3\right)}\right)dr}} \\
& \mkern-18mu \text{considering } q = 1/r^3,\,\, b = \omega\,t_m\,(1-3\,\cos{\theta}^2) \\
& \mkern-18mu \text{as } r\to r_{min},\,q\to q_{min} \text{ and as } r\to \infty,\,q\to 0 \\
&= 1-\frac{2\,\pi}{3\,V}\,\big\langle \frac{1}{q_{min}} + \frac{\cos{\left(b\,q_{min}\right)}}{q_{min}} + b\,Si\left(b\,q_{min}\right)\big\rangle \\
&= 1-\frac{2\,\pi\,\epsilon\,C}{3\,N}\,\bigg[\frac{-2+\Phi(\omega\,t_m\,q_{min})}{q_{min}}\,+\,I_{1,2}\bigg]
\end{split}
\label{eqn:integral_1BB}
\end{equation}
$I_12$ and $\Phi$ are defined as
\begin{equation}\nonumber
\begin{split}
&I_{1,2} = \left\langle b\,Si(b\,q_{min}) \right\rangle_\theta \\
&\Phi(x) = \sqrt{\frac{2\,\pi}{3\,x}}\,\Big(\cos{(x)}\,fc(\sqrt{\frac{6\,x}{\pi}})\,+\,\sin{(x)}\,fs(\sqrt{\frac{6\,x}{\pi}})\Big)
\end{split}
\end{equation}
where $Si(x)$ is the sine integral, $C = N/V$ is given in the unit of molecules per cubic meter. $I_{1,2}$ cannot be evaluated readily and therefore, we expand $Si(x)$ as follows \cite{havil_gamma}
\begin{equation}
Si(x) = \sum_{k=1}^{\infty}{(-1)^{2\,k-1}\,\frac{x^{2\,k-1}}{(2\,k-1)\,(2\,k-1)!}}
\label{eqn:sc_int_exp}
\end{equation}
Using \eref{eqn:sc_int_exp}, $I_{1,2}$ becomes
\begin{equation}
I_{1,2} = -\sum_{k=1}^{\infty}\frac{2\,(-1)^{k}\,q_{min}^{2\,k-1}\,(\omega\,t)^{2\,k}\,_2F_1(\frac{1}{2},-2 k,\frac{3}{2},3)}{(2\,k-1)\,(2\,k-1)!}
\label{eqn:si_int}
\end{equation}
where the hypergeometric function, $_2F_1$ is given by \cite{wolfram_hypergeo}
\begin{equation}\nonumber
\begin{split}
_2F_1(x_1,x_2,x_3,x_4) &= \sum_{n=0}^\infty{(x_1)_n\,(x_2)_n\,x_4^n/(x_3)_n}
\\ &= 1\,+\,\frac{x_1\,x_2\,x_4}{x_3\,1!}\, \\
& +\,\frac{x_1\,(x_1+1)\,x_2\,(x_2+1)\,x_4^2}{x_3\,(x_3+1)\,2!} \\
\end{split}
\end{equation}
The summation in \eref{eqn:si_int} converges for a finite value of \emph{k} and we set its value to 500 for the simulations presented in this work. In this general case, $I_2$ yields
\begin{equation}
\begin{split}
I_2 &= 1-\big\langle \left(1-\cos{\left(a\,t_\xi\right)}\,-\,\cos{\left(a\,t_m\right)}\right)/2 \big\rangle_V \\
& \mkern-18mu \text{considering } q = 1/r^3,\,\, b_x = \omega\,t_x\,(1-3\,\cos{\theta}^2) \\
&= 1-\frac{2\,\pi\,\epsilon\,C}{6\,N}\,\bigg[-\frac{2}{q_{min}}\,+\,\frac{\Phi(b_m\,q_{min})+\Phi(b_\xi\,q_{min})}{q_{min}} \\
&\,+\,\left\langle b_m\,Si(b_m\,q_{min}) \right\rangle\,+\,\left\langle b_\xi\,Si(b_\xi\,q_{min}) \right\rangle\bigg] \\
\end{split}
\label{eqn:integral_2B}
\end{equation}
For brevity, we write the general DQC form factors as
\begin{equation}
\begin{split}
V_{ir} =& V_0\,\exp{\left(-\,I_1^\prime\right)} \\
V_{it} =& -(i/8)\,\big[\exp{\left(-\,I_1^\prime\right)}\,-\, \exp{\left(-\,I_2^\prime\right)}\big] \\
\end{split}
\label{eqn:ndqc_tot_gen}
\end{equation}
where $I_\alpha^\prime = \lim_{N\to\infty}I_\alpha$.
\section*{Results \& Discussion}
\begin{figure}[h]
\begin{center}
    \includegraphics[width=0.75\linewidth]{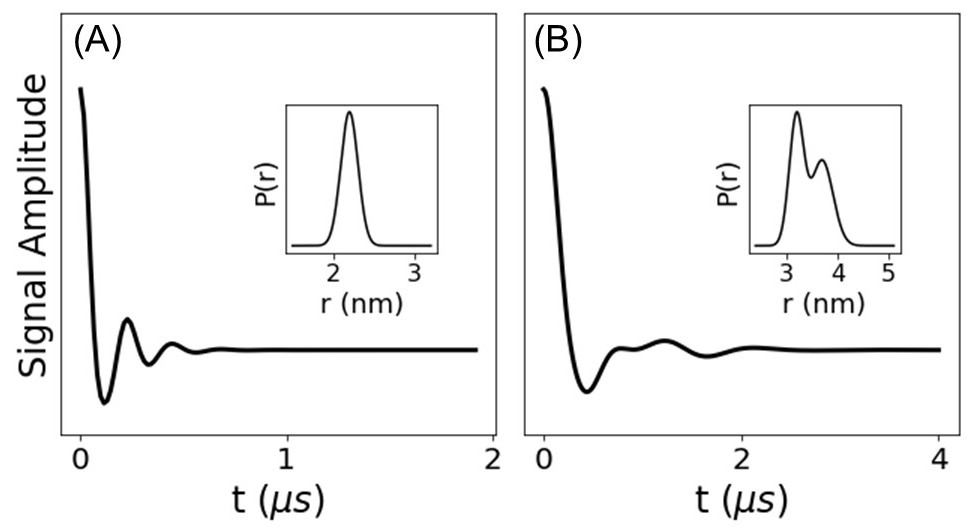}
    \caption{A set of simulated intramolecular DQC signals and the distance distributions (insets) are shown. $t_m$ and the time increments are set to (A) 1.92 $\mu{s}$, 16 ns, and (B) 4.0 $\mu{s}$, 16 ns, with $\delta = 0.8$.}
  \label{fig:pair_examples}
\end{center}
\end{figure}

Usually, the spin pair distance follow a distribution, $P(r)$ in a spin labeled protein and $V_0$, it its discrete form is given by
\begin{equation}
V_0 = \sum_r\,\kappa(r, t)\,P(r)
\end{equation}
where $\kappa(r,t)$ is the dipolar pair kernel. The two simulated DQC pair signals shown in \fref{fig:pair_examples} are used in all the simulations presented in this work.
\begin{figure}[h]
\begin{center}
    \includegraphics[width=0.85\linewidth]{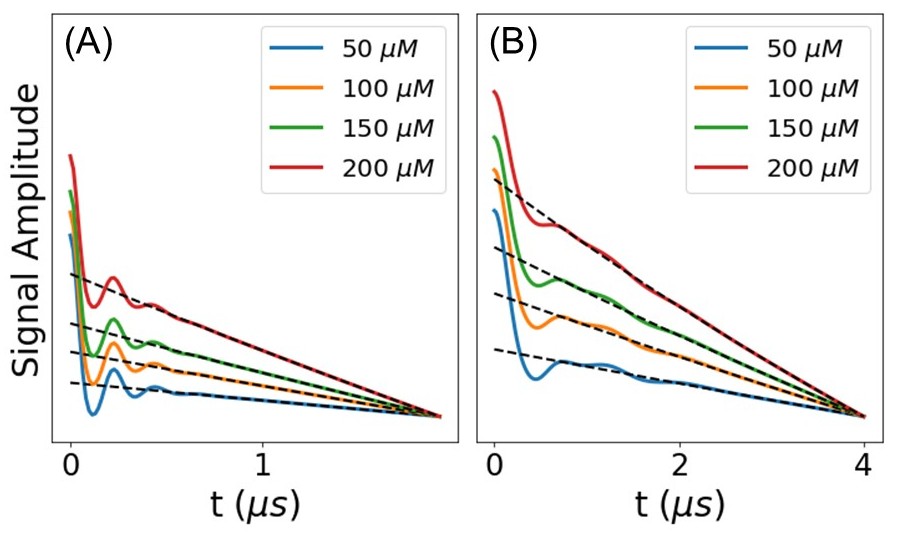}
    \caption{Simulations of DQC time domain signals with varying concentrations (50-200 $\mu{M}$) are shown. The intramolecular signals are taken from \fref{fig:pair_examples} and the background contributions are plotted in black against a common y-axis.}
  \label{fig:dqc_comp1}
\end{center}
\end{figure}
 
\begin{figure}[h]
\begin{center}
    \includegraphics[width=0.6\linewidth]{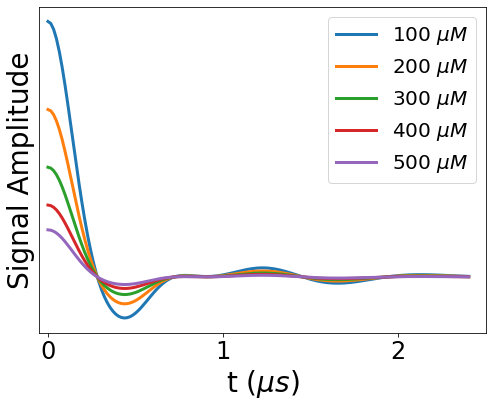}
    \caption{Effect of instantaneous diffusion on the intramolecular DQC signals are shown over a concentration range (100-500 $\mu{M}$). The value of $\delta$ is set to 0.9 and the intramolecular signal corresponds to (B) in \fref{fig:pair_examples}. The plot is truncated at 2.5 $\mu{s}$ for presentation purposes only.}
  \label{fig:dqc_id}
\end{center}
\end{figure}
\begin{figure}[h]
\begin{center}
    \includegraphics[width=0.75\linewidth]{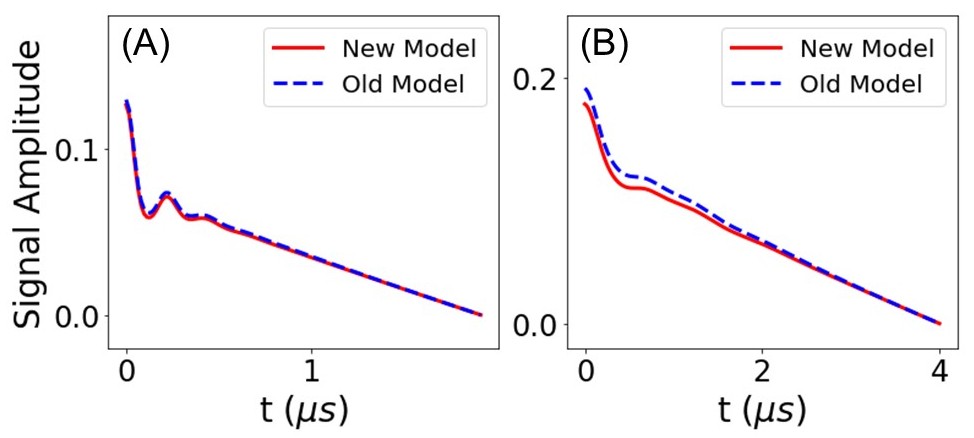}
    \caption{A set of DQC signals are simulated using the old (dotted blue), and the new models (solid red) at a sample concentration of 150 $\mu{M}$. The pair signals are taken from \fref{fig:pair_examples}.}
  \label{fig:model_comp}
\end{center}
\end{figure}
\ypar In \fref{fig:dqc_comp1}, we have presented a set of simulations with the protein concentration varying between 50 and 200 $\mu{M}$. Please note that by protein concentration, we specify the \emph{spin labeled} protein concentration. It is worth mentioning that other than the intermolecular decay, instantaneous diffusion is also responsible for the rapid reduction in signal modulation depth with both increasing concentration and the dipolar evolution time, $t_m$ \cite{many_body_acert,ESR_Methods,ID_decay}. Shown in \fref{fig:dqc_comp1} and \fref{fig:dqc_id} are a set of simulated total, and background corrected DQC signals with increasing concentration. It can be seen that beyond 200 $\mu{M}$, the intermolecular signal amplitude increases rapidly, especially in the pair (B) with $t_m = 4.0 \mu{s}$, while the intramolecular signal amplitude decreases substantially due to the instantaneous diffusion. Hence, the sample concentration in DQC experiments should be kept below 200 $\mu{M}$, especially when longer evolution times are used, posing a challenge in achieving high signal-to-noise ratio (snr) in those cases.
\begin{figure}[!htb]
\begin{center}
    \includegraphics[width=0.9\linewidth]{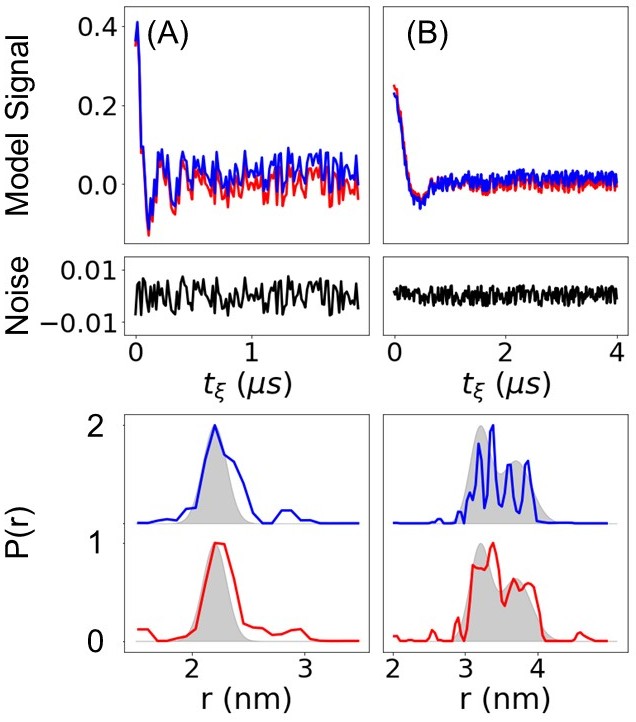}
    \caption{Background corrections (top row) of the signals simulated by the new model in \fref{fig:model_comp} with some added random noise (middle row) by the new (red), and the old methods(blue) are presented. The derived distance distributions are shown in the bottom row along with the actual P(r) (gray shaded area). Note that the derived P(r) in the bottom row are vertically shifted to improve visualization.}
  \label{fig:model_comp2}
\end{center}
\end{figure}
\begin{figure}[!htb]
\begin{center}
    \includegraphics[width=0.8\linewidth]{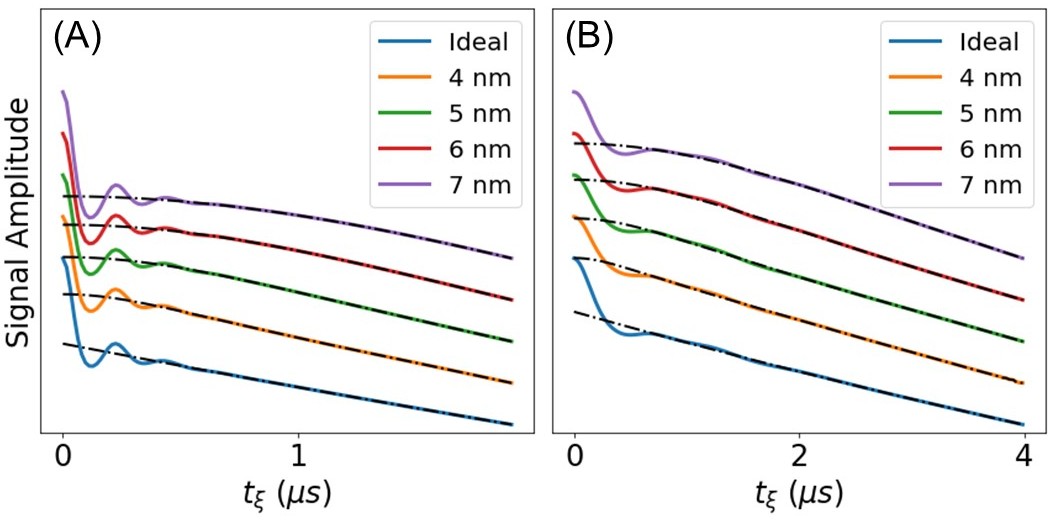}
    \caption{Variation of DQC background signal at a sample concentration of 200 $\mu{M}$ with increasing average minimum distance between two solute molecules in the sample. Note that the signals are vertically shifted to improve visualization.}
  \label{fig:dqc_finsize}
\end{center}
\end{figure}
\xpar We like to compare the effect of using the \emph{new model} presented in this work, and the existing or the \emph{old model} in DQC background corrections, and deriving the distance distributions. The old model is given by \cite{DQC_Theo1}
\begin{equation}
\begin{split}
V_{old} =& V_0\,\exp{(-\epsilon\,\eta_s\,t_m)} \\
& + (1/2)\,\big[-\exp{(\epsilon\,\eta_s\,t_m)}+\exp{(-\epsilon\,\eta_s\,t_\xi)}\big] \\
\label{eqn:old_model}
\end{split}
\end{equation}
where $\eta_s = 2\eta$, considering spin concentration, $C_s$ in the expression rather than spin labeled protein concentration. Note that the DQC form factor derived in this work has the identical intramolecular contribution to that of \eref{eqn:old_model}, and the intermolecular contribution differs only slightly. Hence, we expect both the models to produce nearly identical results at low concentrations. However, at high concentrations and/or for higher values of $t_m$, the results may differ significantly. We simulate a set of DQC signals at 150 $\mu{M}$ concentration of the spin labeled proteins, using the parameters in \fref{fig:pair_examples} by both the methods. The resulting signals are shown in \fref{fig:dqc_comp1} and the difference between the results obtained by the new, and the old models deem rather small.
\ypar Considering that the new model represents the correct form factor of DQC, the simulations in \fref{fig:model_comp} calculated by the new model, added to some random noise are set as the reference. The background correction is done using both the models and a new method for the derivation of distance distribution is used \cite{roy_expds} to demonstrate the effect of background correction on the accuracy of the signal analysis process. It can be seen in \fref{fig:model_comp2} that the background correction by the old model results into significant deviation in the derived distance distribution from that of the actual P(r) and the error amplifies with increasing $t_m$. Therefore, it is necessary to use the correct form factor, even though the two models produce almost similar numerical results.
\xpar Lastly, we demonstrated the effect of finite size of the protein molecules in the sample on the signal and for that, we use the general expression of the signal derived in \eref{eqn:ndqc_tot_gen}. Note that the integrals are evaluated by expanding the sine integral function into a series and as a result of that, the value of $r_min$ must be set to 4 $nm$ or higher. However, it does not pose a major limitation, given that the average radius of a protein molecule is usually higher than the limit. To emphasize the effect, we have shown a set of DQC simulations at relatively high sample concentration of 200 $\mu{M}$ against an increasing $r_{min}$ in \fref{fig:dqc_finsize}. It should be noted that the DQC background signal is small when the concentration and/or $t_m$ are not very large and in such cases, for example (A) in \fref{fig:dqc_finsize}, the effect of $r_{min}$ on the background contribution is not dramatic in the range of 4-7 $nm$. However, its effect is readily visible in case of (B) and a simple linear background subtraction in such cases are likely to affect the derived distance distribution significantly.
\section*{Conclusion}
We have derived the analytical expression and the functional form of the 6-pulse DQC ESR signal in this work. The new, and the previously proposed functional forms vary slightly at low sample concentrations (<50 $\mu{M}$) and for small $t_m$ (<2 $\mu{s}$). However, the difference between the two amplifies at higher concentrations, affecting the accuracy of deriving the distance distributions, especially by non-regularized signal reconstruction methods. The spin labeled proteins are large molecules and with the increasing size of proteins, the DQC background signal shape is likely to shift from linear. The new general functional form of the signal account for the factor. It should be note that the effect of the inefficient inversion pulse is included in a simple, but effective fashion, making the model more generalizable. Additionally, it is possible to utilize the $N$-spin analytical expression of DQC signal to quantify the inhomogeneity in the sample distribution, and multi-spin effects from the experimental time traces.
\subsection*{Acknowledgement}
We sincerely thank Professor John Marohn for providing the Mathematica packages for performing analytical rotation of S=1/2 operators.
\vspace{1cm}
\bibliography{bg_removal}

\end{document}